\documentclass[journal]{IEEEtran}
\usepackage[utf8]{inputenc}
\usepackage{cite}
\usepackage{graphicx}
\usepackage{amsfonts}  
\usepackage{amsmath}
\usepackage{amsthm} 
\newtheorem{definition}{Definition}
\usepackage{algorithm}
\usepackage{algpseudocode}
\algrenewcommand\algorithmicrequire{\textbf{Input:}}
\algrenewcommand\algorithmicensure{\textbf{Output:}}
\usepackage{array}
\usepackage{url}
\begin{document}

\title{On The Detection of Minimum Forecast Horizon For Real-Time Scheduling of Energy Storage Systems in Smart Grid}
\author{Nicholas~Tetteh~Ofoe,~\IEEEmembership{Member,~IEEE,}
        Weilun~Wang,~\IEEEmembership{Member,~IEEE,}
        and~Lei~Wu,~\IEEEmembership{Fellow,~IEEE}%
\thanks{N. Ofoe, W. Wang, and L. Wu are with the Department
of Electrical and Computer Engineering, Stevens Institute of Technology, Hoboken,
NJ, 07030 USA. e-mail: nofoe, wwang110, lwu11@stevens.edu}}%


\maketitle

\begin{abstract}
The increasing integration of energy storage systems (ESSs) into power grids has necessitated effective real-time control strategies under uncertain and volatile electricity prices. An important problem of model predictive control of ESSs is identifying the minimum forecast horizon needed to exactly simulate the globally optimal control trajectory. Existing methods in the literature provide only sufficient conditions and might ignore real-world inconsistencies in control actions. In this paper, we introduce a trajectory-alignment-based definition of the minimum forecast horizon and propose an algorithm that identifies the minimum planning horizon for which all rolling-horizon control decisions match those of the full-horizon global optimization. Using real price data from the bidding zone DK1 in Denmark of the Nord Pool day-ahead market and a realistic ESS model, we illustrate that $60$ hours of forecast horizon allows us to exactly simulate the global control sequence and economic outcomes. In addition, we illustrate that under other parameter configurations, no forecast horizon ensures full convergence, demonstrating the sensitivity of the existence of a forecast horizon to various parameters. Our findings provide an operationally significant framework for minimum forecast horizon detection in storage scheduling and pave the way for the analytical description of this important planning measure.
\end{abstract}


\IEEEpeerreviewmaketitle

\section{Introduction}
\IEEEPARstart{T}{he} decarbonisation of power systems has promoted the rollout of energy storage systems (ESSs) as primary flexibility enablers to stabilize the variability and uncertainty of renewables. Different from traditional generators, ESSs are controlled by energy-constrained dynamics, charging and discharging losses, state of charge (SoC) limitations, and in some instances, self-discharge, representing inherent hurdles to their optimal operation within electricity markets \cite{Schubert9040211}. Successfully leveraging ESSs for economic and operational planning needs real-time control measures that are capable of responding to fluctuating electricity prices in light of long-term operational targets.

For this purpose, infinite-horizon dynamic programming is theoretically optimal, since it incorporates the entire future value of choices. However, the exact solution of these problems is computationally intractable. Consequently, in practice, the most common approaches have been finite-horizon approximations or rolling horizon techniques of planning \cite{Jesudasan6849305} \cite{Cuisinier324245}. Yet, the length of the planning horizon required to approximate infinite-horizon optimality is a key question both in theory and practice.

The works \cite{Garcia2000304} and \cite{Cheevaprawatdomrong2342} originally established the theory of forecast horizons and solution horizons for nonstationary Markov decision processes (MDPs). The basic idea is that one could, under certain conditions of monotonicity and continuity, identify a finite planning horizon for which gathering more information has negligible value to the control policy. Theoretical concepts found their way to applications in planning problems in industries, inventory control problems, and asset-selling problems later on \cite{Bardhan23412}.

Rolling horizon approaches have been implemented in energy systems to manage operations across multiple timescales, for example, day-ahead vs. intraday scheduling for storage \cite{finnah2022scheduling} or power system ramping services \cite{Ela201555}. Most, nonetheless, are based on heuristically fixed horizons (e.g., 24 or 48 hours), for which optimality or suboptimality is generally not formally guaranteed. Such informality may generate inefficient scheduling or unforeseen market activity, particularly in the case of extreme price volatility and/or physical limitations.

A recent influential work in \cite{prat2024longlongenoughfinitehorizon} revisited the forecast horizon issue in the framework of deterministic storage operation. The authors provided a mathematical condition, stating that if the two constrained instances of the finite-horizon problem have the same SoC values at the end of a decision-making interval, the planning horizon is a good forecast horizon. The authors also provided an algorithm for determining the minimum forecast horizon and a suboptimality bound. The authors' findings were substantiated with case studies employing data for Danish electricity prices.

Nonetheless, the stated sufficient and necessary condition in \cite{prat2024longlongenoughfinitehorizon} may not be the case in reality. Because of the possible non-uniqueness of the optimal solutions to ESS scheduling problems, the fact that the SoC values at the decision horizon are not equal does not immediately imply that a planning horizon $T$ is not a forecast horizon. As mentioned earlier in \cite{Garcia2000304}, sufficient conditions cannot eliminate the possibility of optimality in the general case. As a consequence, the algorithm in \cite{prat2024longlongenoughfinitehorizon} might overestimate the forecast horizon or miss valid ones in the presence of many different terminal state equivalent trajectories.

This paper intends to revisit the notion of forecast horizons based on an action-oriented framework, and introduce a modified method to define forecast horizons based on trajectory equivalence instead of SoC endpoint matching. We construct a simulation framework in which a global (i.e., a sufficiently long horizon) benchmark is contrasted with rolling forecasts of different lengths to ascertain the most compact planning horizon that also mimics the optimal control sequence. We apply our method to a stylized ESS model with realistic efficiency costs and price volatility. By doing so, we illustrate that forecast horizons may both be shorter and more strongly identified than the current sufficient condition in \cite{prat2024longlongenoughfinitehorizon}.

The rest of the paper is organized as follows: ****

 

\section{Methodology}
\subsection{Problem Formulation and Minimum Forecast Horizon}

We consider the issue of optimally scheduling a price-taking ESS for a finite planning horizon \( \mathcal{T} = \{1, 2, \dots, T\} \), based on available price forecasts. The ESS is characterized by energy capacity, charging power, and discharging power bounds, conversion loss efficiencies, and storage deterioration through leakage. Our aim is to identify the minimum planning horizon \( T^* \) for which the first-stage control decision under rolling optimization is consistent with the global infinite-horizon optimal decision, a definition which we pose as the minimum forecast horizon.

Let \( s_t \) denote the SoC at time \( t \), and \( p_t^c \) and \( p_t^d \) denote the charging and discharging powers, respectively. Only one of \( p_t^c \) or \( p_t^d \) can be nonzero at any time. The model aims to maximize arbitrage profit based on a vector of given electricity prices \( \mathcal{C} \in \mathbb{R}_+^{T} \).


We use the standard deterministic ESS model that was presented in \cite{prat2024longlongenoughfinitehorizon} to build our optimization model \( \mathcal{S}(\mathcal{T}, \mathcal{C}) \) as follows:

\begin{subequations}\label{eq:mainproblem}
\begin{align}
\max_{p_t^c,\, p_t^d,\, s_t} \sum_{t \in \mathcal{T}} &\Delta t \cdot C_t  \cdot  (p_t^d - p_t^c) \label{eq:mainproblem_obj} \\[2ex]
\text{s.t. } \, 
s_1 = \rho \cdot &s^{\text{init}} + \Delta t \cdot \left(\eta^c \cdot p_1^c - \frac{1}{\eta^d} \cdot p_1^d\right), \quad t = 1 \label{eq:mainproblem_init} \\[2ex]
s_t = \rho \cdot &s_{t-1} + \Delta t \cdot \left(\eta^c \cdot p_t^c - \frac{1}{\eta^d} \cdot p_t^d\right), \, \, \, \forall t \in \mathcal{T} \setminus \{1\}, \label{eq:mainproblem_dyn} \\[2ex] 
\underline{s} &\leq s_t \leq \overline{s}, \quad \quad \quad \quad \quad \quad \quad \quad \quad \quad \forall t \in \mathcal{T}, \label{eq:mainproblem_socbounds} \\[2ex]  
0 &\leq p_t^c \leq \overline{P}^c, \quad \quad \quad \quad \quad \quad \quad \quad \quad \forall t \in \mathcal{T}, \label{eq:mainproblem_chargebounds} \\[2ex] 
0 &\leq p_t^d \leq \overline{P}^d, \, \, \, \, \, \quad \quad \quad \quad \quad \quad \quad \quad \forall t \in \mathcal{T}, \label{eq:mainproblem_dischargebounds} \\[2ex] 
&p_t^c \cdot p_t^d = 0, \, \quad \quad \quad \quad \quad \quad \quad \quad \quad \forall t \in \mathcal{T}, \label{eq:mainproblem_binary}
\end{align}
\end{subequations}    


Here, \( \eta^c \in (0,1] \) and \( \eta^d \in (0, 1] \) are the charging and discharging efficiencies; \( \rho \in (0, 1] \) captures leakage; $\overline{P}^c$ and $\overline{P}^d$ are the maximum charging and discharging power respectively; \( \Delta t \) is the time discretization step. The exclusive charge and discharge constraint~\eqref{eq:mainproblem_binary} can be enforced via a binary control variable in the implementation to prevent the simultaneous charging and discharging of ESSs.

Our approach differs from that in \cite{prat2024longlongenoughfinitehorizon} by not focusing on matching final SoC values \( \underline{s}_H = \overline{s}_H \) where $H$ is the last time slot of the forecast horizon, but instead on matching \textit{first-stage actions} \( p_1^* \) between rolling horizon and global optimization.

\begin{definition}[Minimum Forecast Horizon]
Let \( \{p_t^*\}_{t=1}^{T_\text{max}} \) denote the optimal control sequence from a global optimization over the full horizon, and \( \hat{p}_t \) denote the first decision from the rolling horizon optimization over \( [t, t+T] \). A planning horizon \( T \) is a \textit{forecast horizon} if \( |\hat{p}_t - p_t^* | \leq \varepsilon \) $\forall t $,. The \textit{minimum forecast horizon} \( T^* \) is the smallest such \( T \).
\end{definition}

Our definition of the minimum forecast horizon avoids the sufficiency-only limitation of the theorem in \cite{finnah2022scheduling}. This definition directly supports real-time control design.

\subsection{Global Optimization Baseline}

The \textit{global optimality baseline} serves to provide a reference against which we measure the efficiency and adequacy of finite horizon forecasts. It is the hypothetically optimal control strategy with full knowledge of the future throughout the entire planning horizon \( \mathcal{T}_{\text{max}} = \{1, \dots, T_{\text{max}}\} \), where \( T_{\text{max}} \) can represent several weeks or perhaps even a few months of operation. It is computationally intensive to solve for but is useful for two important reasons: benchmarking accuracy and detecting forecast horizons, since it offers the ground truth sequence \( \{p_t^*\} \) against which the solutions of rolling horizon optimizations are compared.

We solve the full-horizon optimization of the deterministic ESS scheduling problem \( \mathcal{S}(\mathcal{T}_{\text{max}}, \mathcal{C}) \), where \( \mathcal{C} = \{C_1, \dots, C_{T_{\text{max}}}\} \) is the vector of given electricity prices. This yields the globally optimal state and control trajectories that meet the system dynamics and operational constraints described in the model (1):
\begin{align}
    \{s_t^*, p_t^{*,c}, p_t^{*,d}, \mathcal{P}_{t}^{*}\}_{t=1}^{T_{\text{max}}}
\end{align}

To ensure robustness, the solution accounts for time-changing prices comprising high-frequency volatility and even price spikes. It also accounts for physical limitations of the ESS, such as limitations on energy, ramping through power limitations, and non-simultaneity. Finally, the solution accounts for round-trip inefficiencies and leakage effects that influence long-run arbitrage profitability.

We stored the globally optimal trajectory in a structured format and used it for the comparison of the first-stage actions produced from rolling horizon optimizations with shorter horizons \( T < T_{\text{max}} \). We can then determine the minimum \( T^* \) with which the rolling horizon solutions replicate the globally optimal control path.

It is noted that the global solution is also an approximation to the infinite-horizon solution \( \mathcal{S}(\mathbb{N}^+, C) \) since the computation of the solution to an infinite problem is impossible. We assume that for large-enough \(T_{\text{max}} \), the marginal value of extending the horizon becomes negligible, a common supposition in approximate dynamic programming and long-horizon energy scheduling problems~\cite{Jesudasan6849305, Garcia2000304}.

Ultimately, this solution gives a quantitative estimate of the \textit{performance loss} due to finite planning horizons, enabling us to effectively calculate measures like aggregate profit shortfall, average SoC deviation, control discrepancies, and suboptimality rates for each candidate horizon \( T \). With these measures, we can justify whether the minimum forecast horizon \( T^* \) is operationally reasonable, and whether additional robustness strategies are warranted in a live environment.

\subsection{Rolling Horizon Framework}

In practice, system operators are unable to solve large-scale global problems with long horizons in real operational settings due to the computational time constraint and forecast unreliability. Rather, they follow \textit{rolling horizon optimization} (RHO) or \textit{model predictive control} (MPC), which at each instant optimizes a short-horizon problem and takes only the first-stage decision. It then moves one time step forward whenever new data becomes available.

The rolling horizon framework approximates the global optimum by iteratively rolling out the local solutions to the planning problem \( \mathcal{S}(\{t, \dots, t+T-1\}, C_t^{(T)}) \), where \( T \) is the planning horizon and \( C_t^{(T)} \) is the local portion of the price vector. Only the first control move \( \hat{p}_t^{(T)} \) is executed at each time step \( t \), and the state is updated thereafter.

The minimum forecast horizon search algorithm using the rolling horizon framework is presented in \textbf{Algorithm 1}, where \( T_{\text{max}} \) is the length of the full planning horizon and \( T \in \{2, \dots, T_{\text{max}}\} \) is the current candidate planning window.

\begin{algorithm}[H]
\caption{Minimum Forecast Horizon Search}
\label{alg:forecast_horizon}
\begin{algorithmic}[1]
\Require Given price vector $C \in \mathbb{R}^{T_{\max}}$, parameters of the ESS, and initial SoC $s^{init}$
\Ensure The minimum forecast horizon $T^*$ such that rolling policy matches global optimal actions

\vspace{1mm}

\State \textbf{Step 1: Solve Global Optimization}
\State Solve $\mathcal{S}(\mathcal{T}_{\max}, C)$ to obtain $\{p_t^*\}_{t=1}^{T_{\max}}$
\vspace{1mm}

\State \textbf{Step 2: Rolling Horizon Optimization}
\For{$T = 2$ to $T_{\max}$}
    \State Initialize $s \gets s^{init}$
    \State Initialize match flag $\texttt{match} \gets \texttt{True}$

    \For{$t = 1$ to $T_{\max} - T + 1$}
        \State Extract price window $C_t^{(T)} \gets \{C_t, \dots, C_{t+T-1}\}$
        \State Solve $\mathcal{S}(\{t,..., t+T-1\}, C_t^{(T)})$ with initial SoC being $s$
        \If{no feasible solution}
            \State $\texttt{match} \gets \texttt{False}$; \textbf{break}
        \EndIf
        \State Let $\hat{p}_t^{(T)}$ be the first-stage control action
        \vspace{1mm}
        \If{$|\hat{p}_t^{(T)} - p_t^*| > \varepsilon$}
            \State $\texttt{match} \gets \texttt{False}$; \textbf{break}
        \EndIf
        \State Update $s \gets$ storage dynamics using $s_t$ from $\hat{p}_t^{(T)}$
    \EndFor

    \If{$\texttt{match} = \texttt{True}$}
        \State \Return $T^* \gets T$
    \EndIf
\EndFor
\State \Return ``No forecast horizon found up to $T_{\max}$''
\end{algorithmic}
\end{algorithm}

For greater computational tractability, the SoC at every stage is propagated forward, maintaining feasibility for time-coupled constraints. This is to circumvent computing solutions of infeasible subproblems from mismatched initial conditions. Moreover, previous solver outputs can be utilized to provide substantial acceleration of convergence for high-resolution control (e.g., hourly or sub-hourly).

In real-world systems with model error, noise, or action non-uniqueness, equality \( \hat{p}_t^{(T)} = p_t^* \) might be inexact. For this case, we consider a numerical tolerance \( \varepsilon > 0 \) and modify the definition of the control matching condition as follows:
\begin{equation}
    \left\| \hat{p}_t^{(T)} - p_t^* \right\| \leq \varepsilon, \quad \forall t
\end{equation}
This is a relaxed condition to accommodate solver tolerances and discretization errors. A smaller \( \varepsilon \) is associated with more stringent convergence demands and generally leads to longer forecast horizons identified. An increase in \( \varepsilon \), however, is likely to produce shorter horizons that are suboptimal and may be acceptable in practice.

The algorithm itself has a number of advantages over the condition-based scheme in \cite{prat2024longlongenoughfinitehorizon} as it tests for action-level equivalence directly instead of SoC convergence, thus providing a stronger alignment with control policy outputs. 

\section{Numerical Results}
We present the experimental design and simulation results for assessing \textbf{Algorithm 1} to identify the minimum forecast horizon. We explain the test system setup, data employed, and measured control performance over different forecast horizon lengths.

\subsection{Simulation Setup}
The simulations take into account a single ESS that is active for 91 days (from January 1 through March 31, 2024) at an hourly resolution, resulting in a total planning horizon of $T_{\text{tot}}=2,184$ hours. The electricity price vector $\mathcal{C} \in \mathbb{R}^{2184}$ was extracted from the entsoe transparency platform day-ahead market for the bidding zone DK1 in Denmark \cite{entsoe2024}. The data show typical market volatility with periodic cycles as well as sudden spikes in prices.

\begin{figure}[!h]
\centering
\includegraphics[width=3.4in]{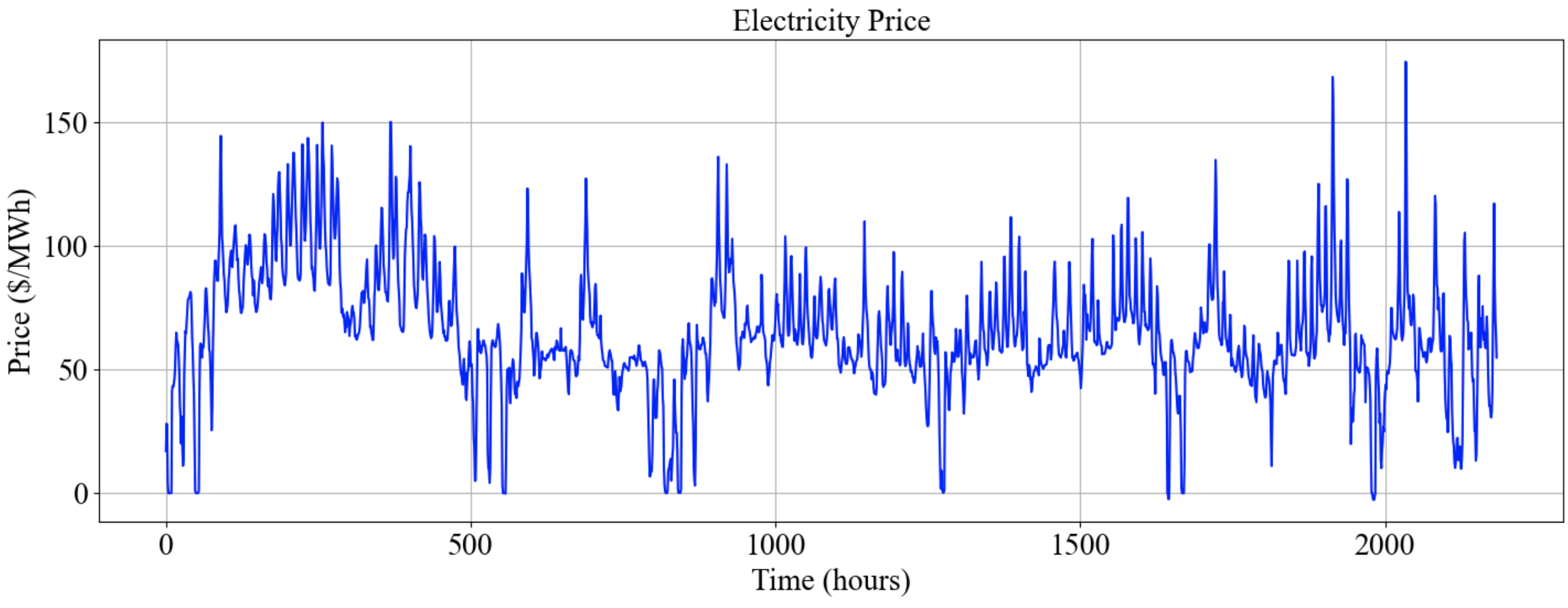}
\caption{Day-ahead hourly electricity prices of DK1 between January and March 2024.}
\label{fig_sim}
\end{figure}

The parameters of the ESS are presented in Table~\ref{storage_params}. A binary control logic is used for prohibiting simultaneous charging and discharging. The simulations were all conducted using Python 3.11 and Gurobi 12.0.2.

\begin{table}[!t]
\renewcommand{\arraystretch}{1.3}
\caption{ESS Parameters}
\label{storage_params}
\centering
\begin{tabular}{|c|c|c|}
\hline
\textbf{Parameter} & \textbf{Description} & \textbf{Value} \\
\hline
$\overline{P}^c$ & Max charging power (MW) & 1.0 \\
\hline
$\overline{P}^d$ & Max discharging power (MW) & 1.0 \\
\hline
$\eta^c$ & Charging efficiency & 0.85 \\
\hline
$\eta^d$ & Discharging efficiency & 0.85 \\
\hline
$E_{\min}$ & Min state of charge (MWh) & 0.0 \\
\hline
$E_{\max}$ & Max state of charge (MWh) & 10.0 \\
\hline
$E_{\text{init}}$ & Initial state of charge (MWh) & 5.0 \\
\hline
$\rho$ & Leakage factor (per hour) & 0.99 \\
\hline
$\Delta t$ & Time step (hours) & 1.0 \\
\hline
$H$ & Decision horizon (steps) & 1 \\
\hline
$T_{\text{tot}}$ & Total simulation duration (hours) & 2,184 \\
\hline
\end{tabular}
\end{table}


We implement \textbf{Algorithm 1} for detecting the minimum forecast horizon over various planning window sizes of $T \in [2,88]$, with each rolling simulation being compared to the optimal global path. For each candidate $T$, an optimal rolling control action at each time step $t$ is computed and the first control action $\hat{p}_t^{(T)}$ is stored. These actions are compared with full-horizon optimal controls $p_t^*$ achieved via a one-shot optimization over the complete planning horizon. The decision comparison is made using a numerical tolerance $\varepsilon =  10^{-4}$. Given that this condition holds for all $t  \in \{ 1, \ldots, T_{\text{tot}} - T + 1 \}$, the corresponding value of $T$ is declared as the minimum forecast horizon $T^*$.

\subsection{Results and Discussion}
\subsubsection{Global Optimal Control Solution}
\begin{figure}[!h]
\centering
\includegraphics[width=3.4in]{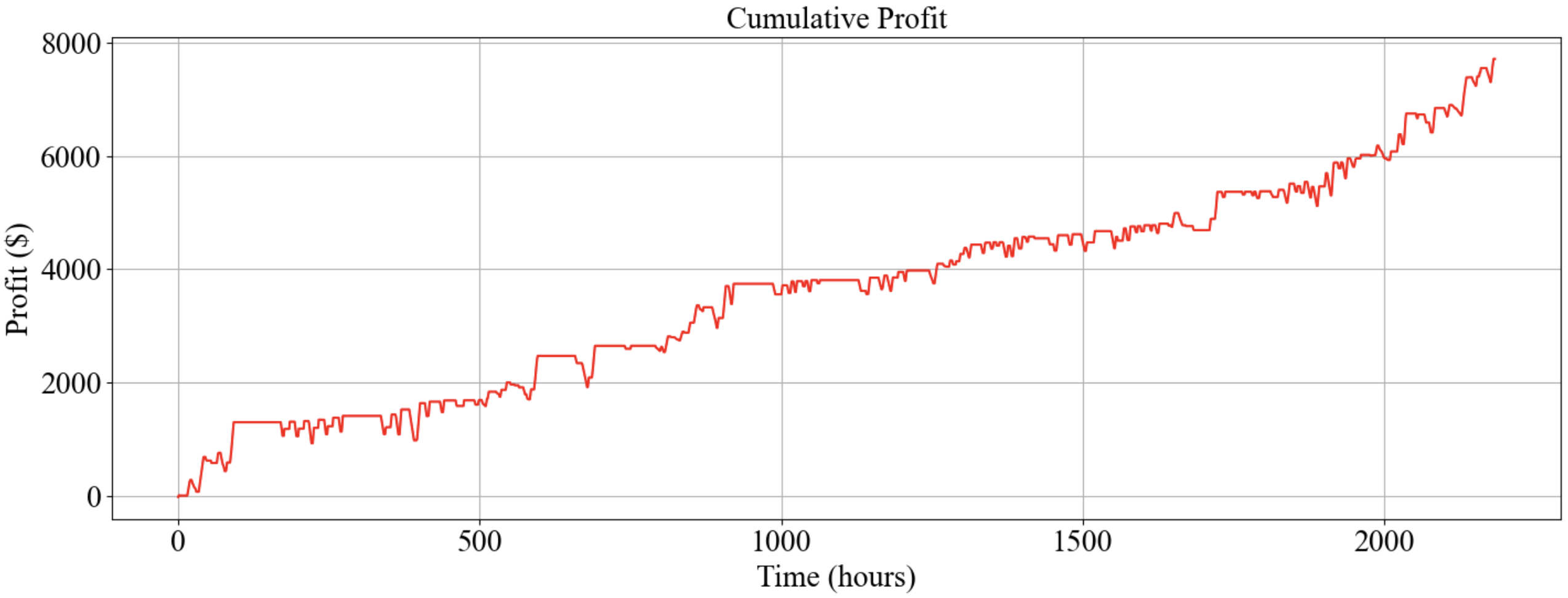}
\caption{Cumulative profit from the global optimization over 2,184 hours.}
\label{fig:global_profit}
\end{figure}

Fig.~\ref{fig:global_profit} shows the accumulated profit from the one-shot optimization throughout the entire 2184-hour simulation period. The ESS accumulates arbitrage over time with significant profitability jumps at times of peak market prices. Piecewise-flat segments over certain periods reveal areas with neither charging nor discharging being profitable and show the impact of round-trip efficiency and price spread.

\begin{figure}[!h]
\centering
\includegraphics[width=3.4in]{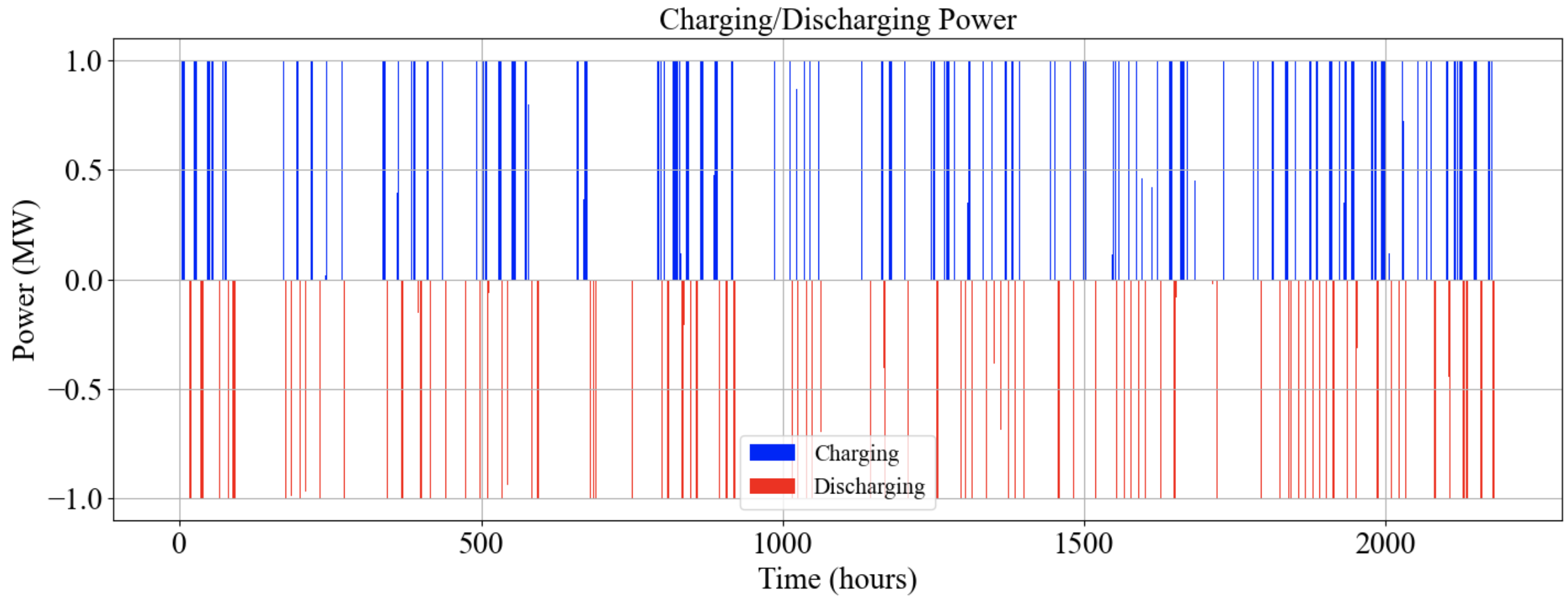}
\caption{Global optimal charging and discharging schedule.}
\label{fig:charge_discharge}
\end{figure}

The charging and discharging power paths under the global optimal policy are shown in Fig.~\ref{fig:charge_discharge}. The blue (charging) and red (discharging) profiles clearly impose mutual exclusivity. Charging is accumulated during low-price periods and is concentrated during high-price periods for discharging. The more frequent switch during volatile periods indicates the responsiveness of the system to steep price changes.

\begin{figure}[!h]
\centering
\includegraphics[width=3.4in]{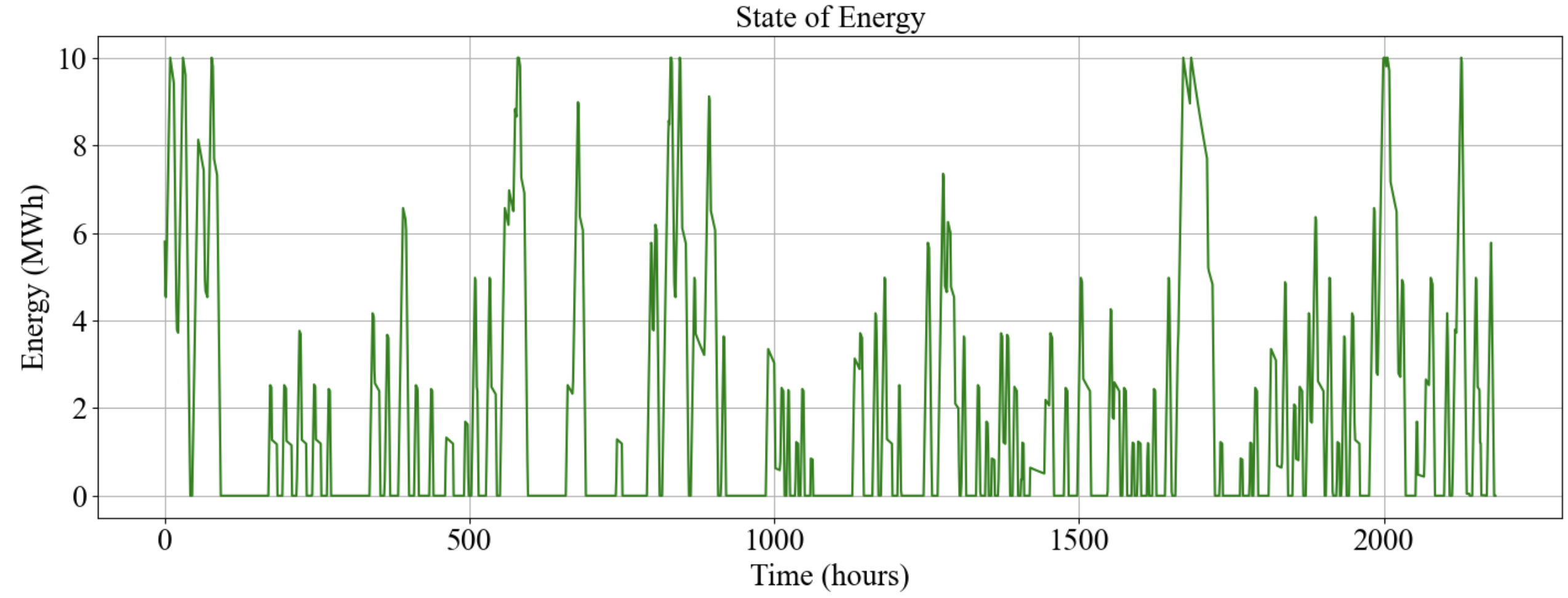}
\caption{State of energy (SoE) trajectory under the global solution.}
\label{fig:soe_trajectory}
\end{figure}

The SoC trajectory is depicted in Fig.~\ref{fig:soe_trajectory}. The trajectory indicates how the ESS prefers to drain or fill up in advance to meet anticipated price trends. Periods with continuous low SoC correspond to periods of net discharge; high SoC spikes appear during periods of low-price recovery. The behavior is indicative of effective energy cycling throughout the planning horizon without breaching capacity limits.

Figs.~\ref{fig:global_profit}-\ref{fig:soe_trajectory} support the claim that the global optimization yields a reference trajectory. We are able to see interpretable responses to market price signals. Thus, the global optimization maintains physical feasibility and maximizes long-term profit.

\subsubsection{Rolling Horizon Convergence and Minimum Forecast Horizon}
\begin{figure}[!h]
\centering
\includegraphics[width=3.4in]{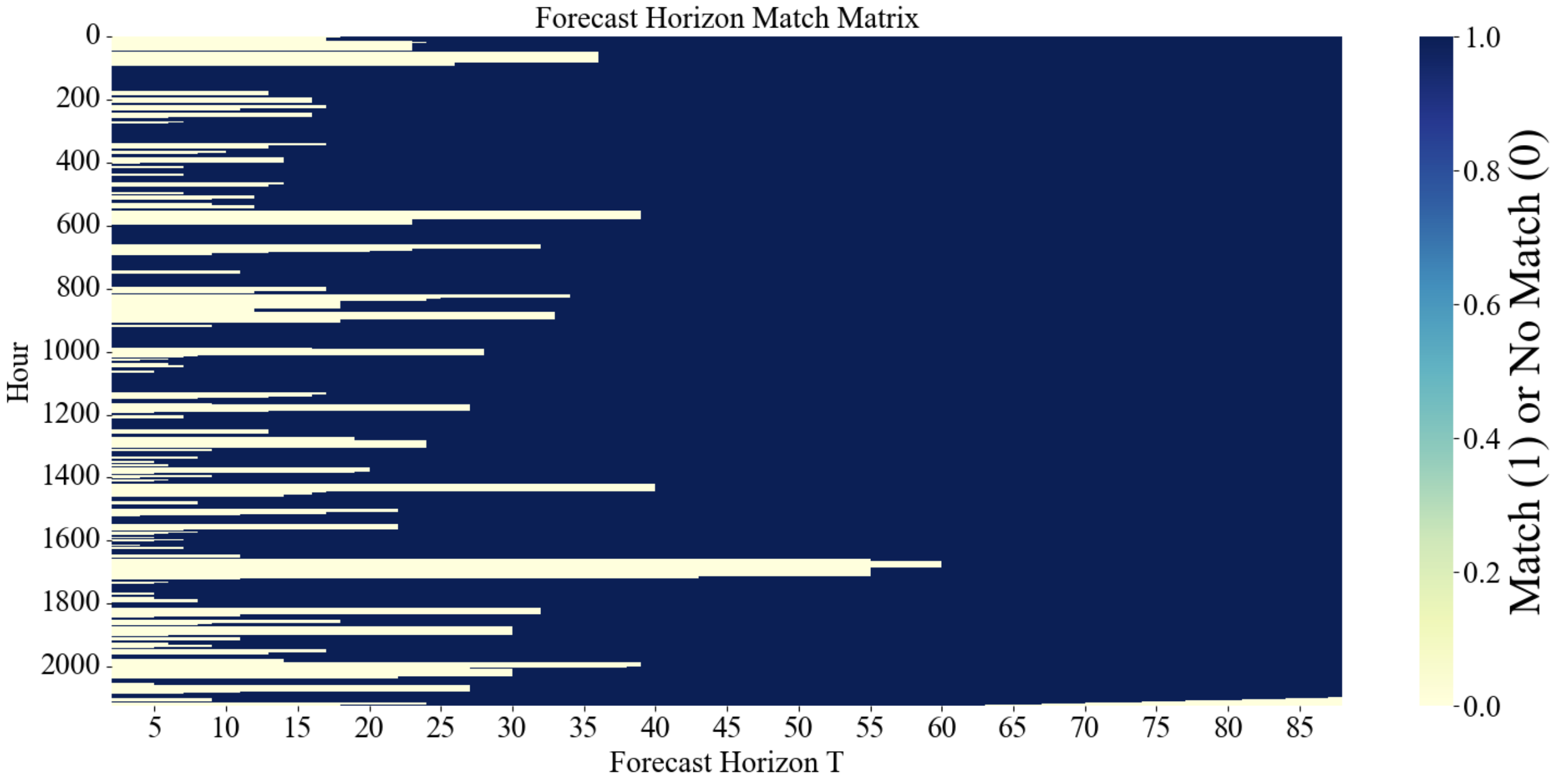}
\caption{Forecast horizon match matrix.}
\label{fig:horizon_match_matrix}
\end{figure}

To determine the minimum forecast horizon $T^*$, we compare the first-stage control decision match between rolling optimization and the global optimum over forecast horizons from $2$ to $88$ hours. Fig.~\ref{fig:horizon_match_matrix} is a binary heatmap representing whether at each time $t$ the rolling decision equaled the global optimum for each horizon $T$.

The blue region is dense and indicates that optimal alignment starts at $T=60$ and hence its identification as the minimum forecast period required. Misalignment (light areas) is more common for small values of $T$, especially over periods with high volatility or high prices. It is therefore imperative to note that the typical planning windows of $24$ or $48$ hours, although commonly used in practice, may be insufficient to preserve global optimality in this case.

\begin{figure}[!h]
\centering
\includegraphics[width=3.4in]{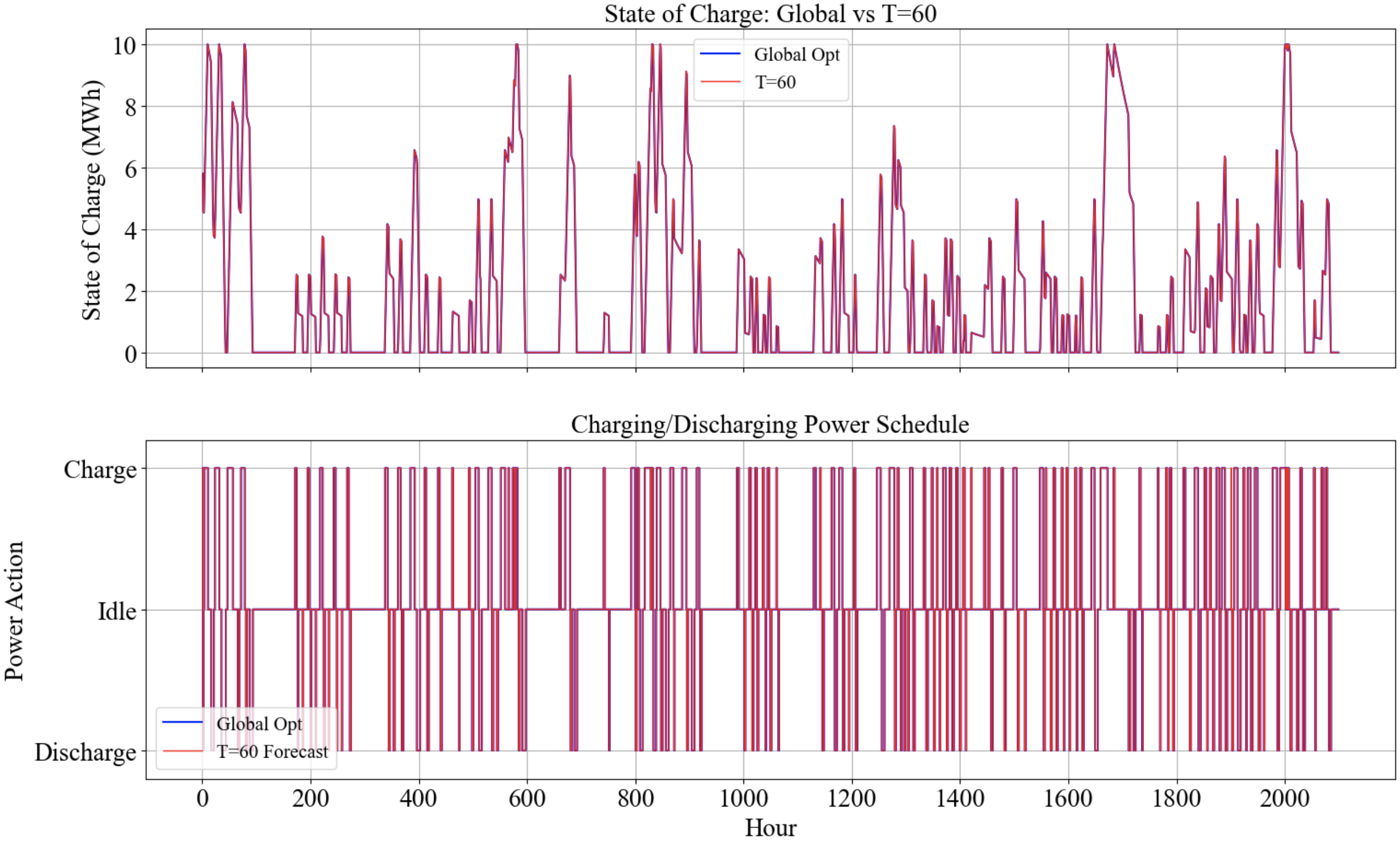}
\caption{Comparison of the global optimal solution and the rolling horizon trajectory with $T = 60$.}
\label{fig:global_vs_rolling_trajectories}
\end{figure}

Fig.~\ref{fig:global_vs_rolling_trajectories} further shows a comparison between SoC (upper part) and power schedule (lower part) of global optimal solution and the rolling horizon solution with $T=60$. The results match almost perfectly, demonstrating that  $T=60$ is enough for replicating both state and action trajectories.

\begin{figure}[!h]
\centering
\includegraphics[width=3.4in]{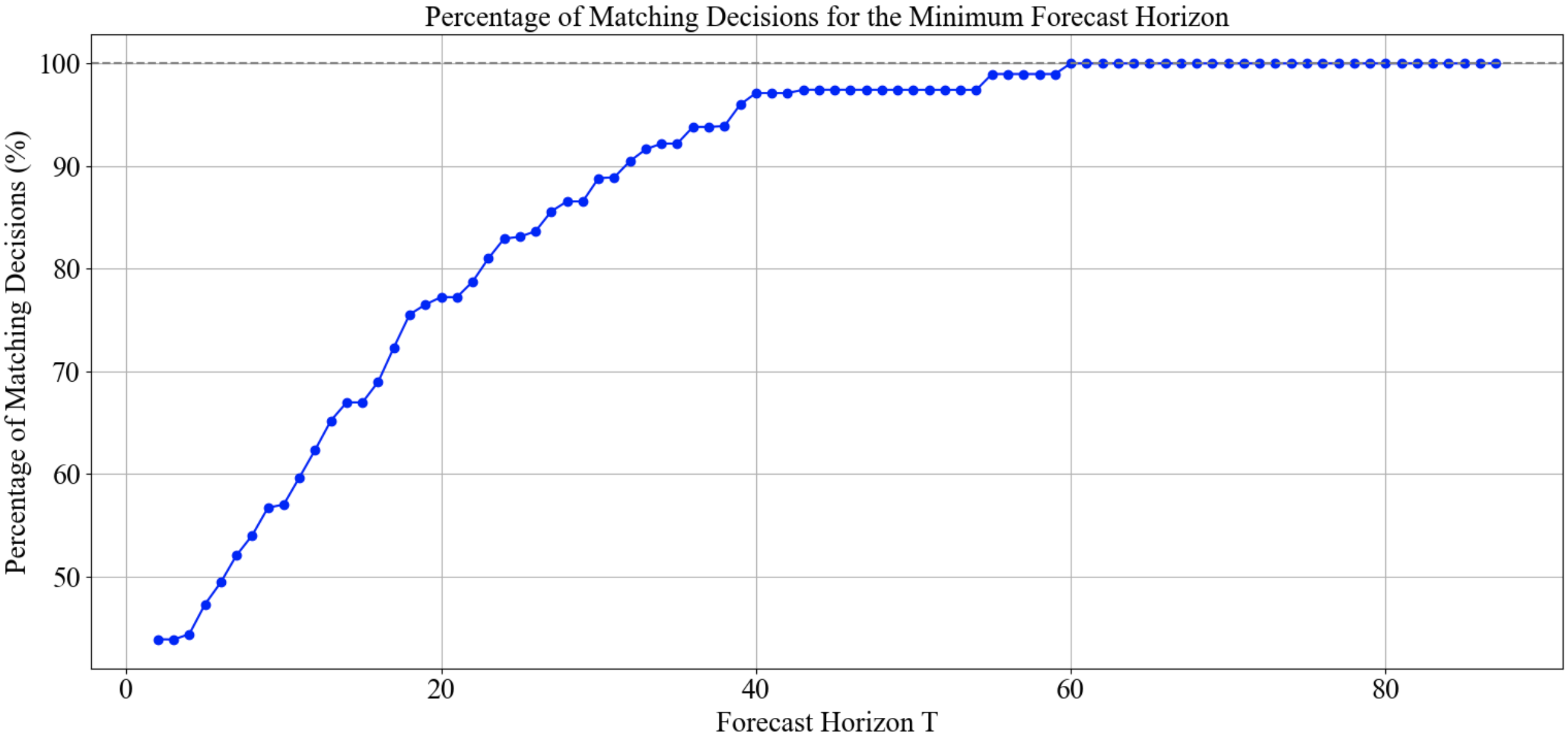}
\caption{Percentage of matching decisions between rolling and global control against different forecast horizons.}
\label{fig:match_percentage_curve}
\end{figure}

Fig.~\ref{fig:match_percentage_curve} shows the percentage match of first-stage decisions plotted as a function of forecast horizon length. The matching increases monotonically to 100\% at $T=60$, when all rolling decisions match the global reference. This gives strong empirical evidence supporting trajectory-based control convergence.

\begin{figure}[!h]
\centering
\includegraphics[width=3.4in]{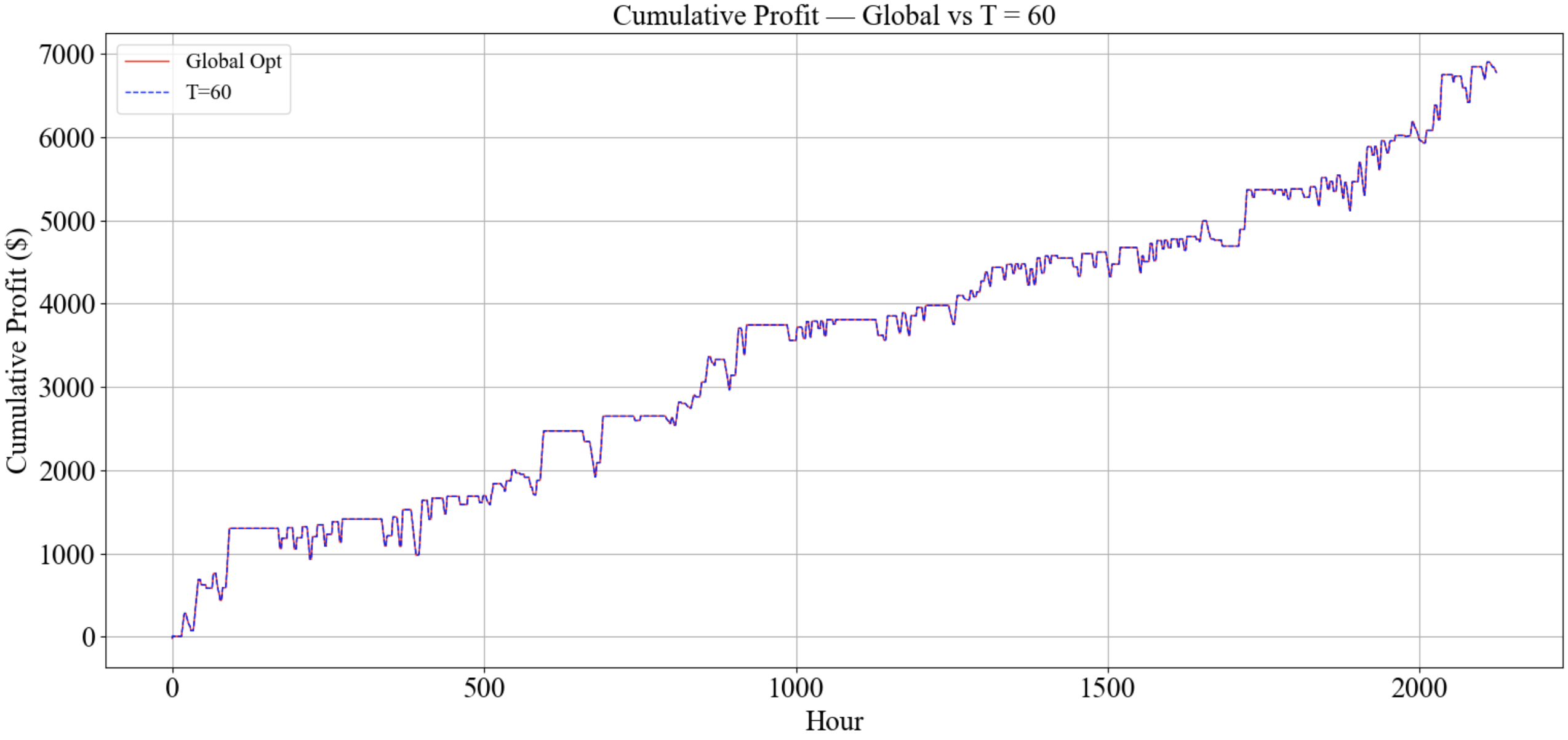}
\caption{Cumulative profit comparison between global optimization and rolling horizon with $T=60$.}
\label{fig:profit_T60}
\end{figure}

In order to confirm that the minimum forecast horizon is economically significant, we compare the total profit of the rolling horizon optimization with $T=60$ to that of the global benchmark. From Fig.~\ref{fig:profit_T60}, one can visually confirm that these two trajectories coincide, verifying that having an exact forecast horizon of  $T^* =60$ is able to exactly imitate not only the optimal control actions, but the total arbitrage profit as well.

\begin{figure}[!h]
\centering
\includegraphics[width=3.4in]{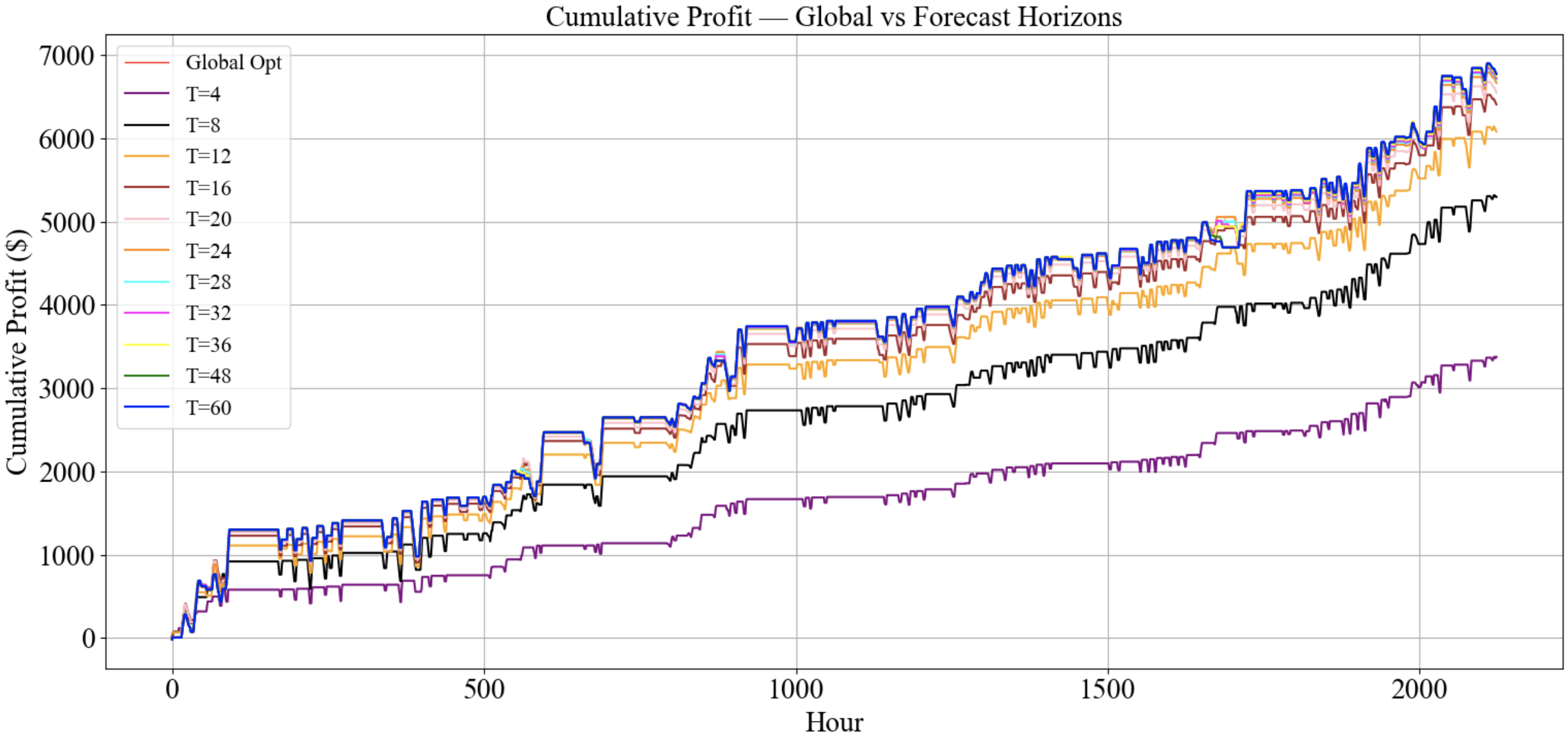}
\caption{Cumulative profit under different forecast horizons.}
\label{fig:profit_allT}
\end{figure}

Fig.~\ref{fig:profit_allT} further shows the cumulative profit for several different forecast horizons (i.e., $T=$4, 8, 12, 16, 20, 24, 28, 32, 36, 48, and 60) against the global solution. We observe that small values of $T$ cause very significant degradation of performance. At $T=4$, profit is well below optimal by more than $40\%$. Improvements continue to be made as $T$ increases, but convergence is not reached until  $T \approx 60$. This confirms the pitfalls of the arbitrary choice of short forecast horizons in real systems.

\subsubsection{Sensitivity of Forecast Horizon to ESS Parameters}
To investigate whether the existence of a minimum forecast horizon depends on the ESS parameters, we run our algorithm with the following changes from Table~\ref{storage_params}: $\eta^c=\eta^d=0.9$ and $\rho=1.0$. 
These parameters align with what was used in \cite{prat2024longlongenoughfinitehorizon}.

\begin{figure}[!h]
\centering
\includegraphics[width=3.4in]{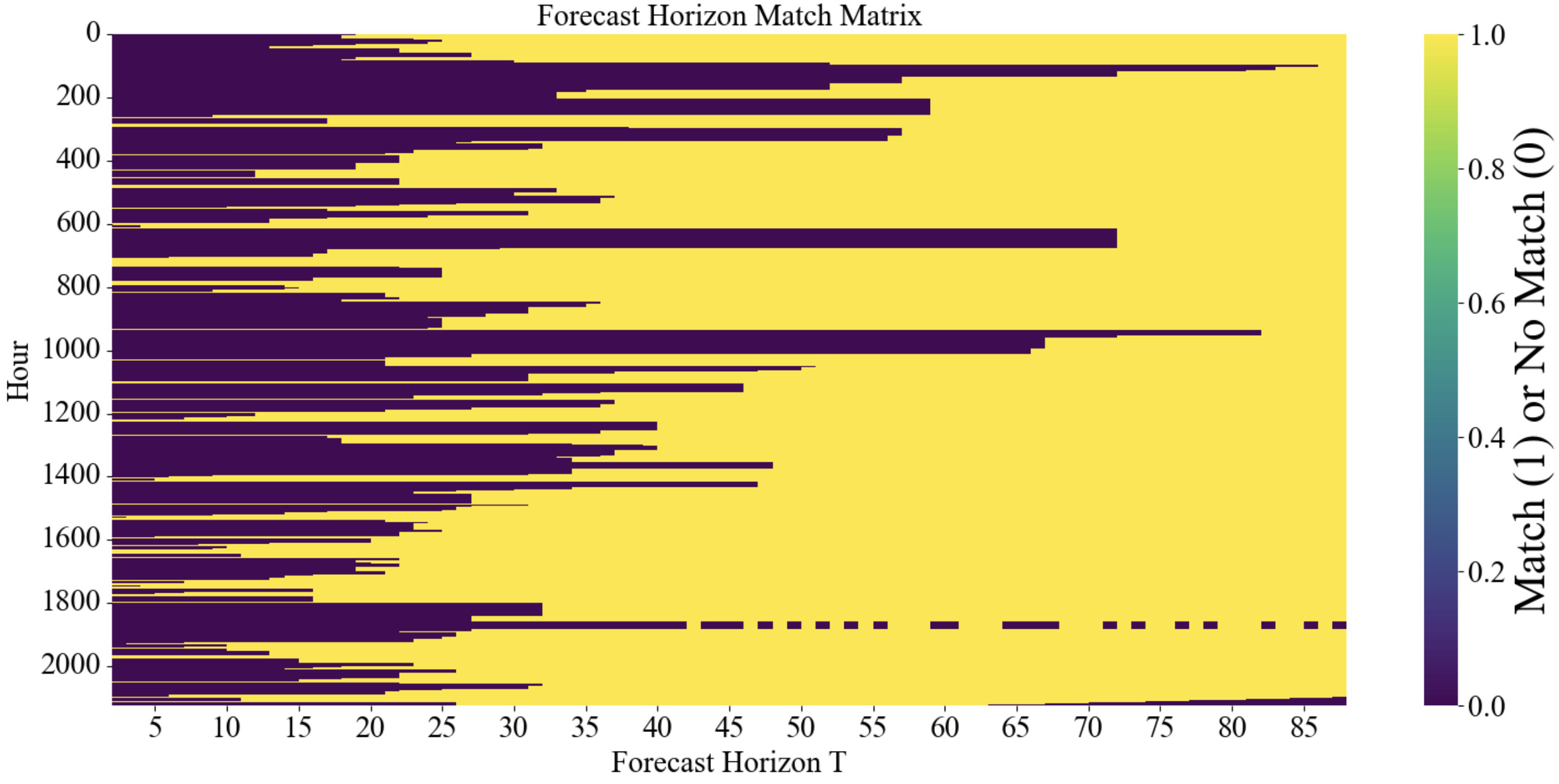}
\caption{Forecast horizon match matrix using the new parameters.}
\label{fig:no_min_horizon_matrix}
\end{figure}
\begin{figure}[!h]
\centering
\includegraphics[width=3.4in]{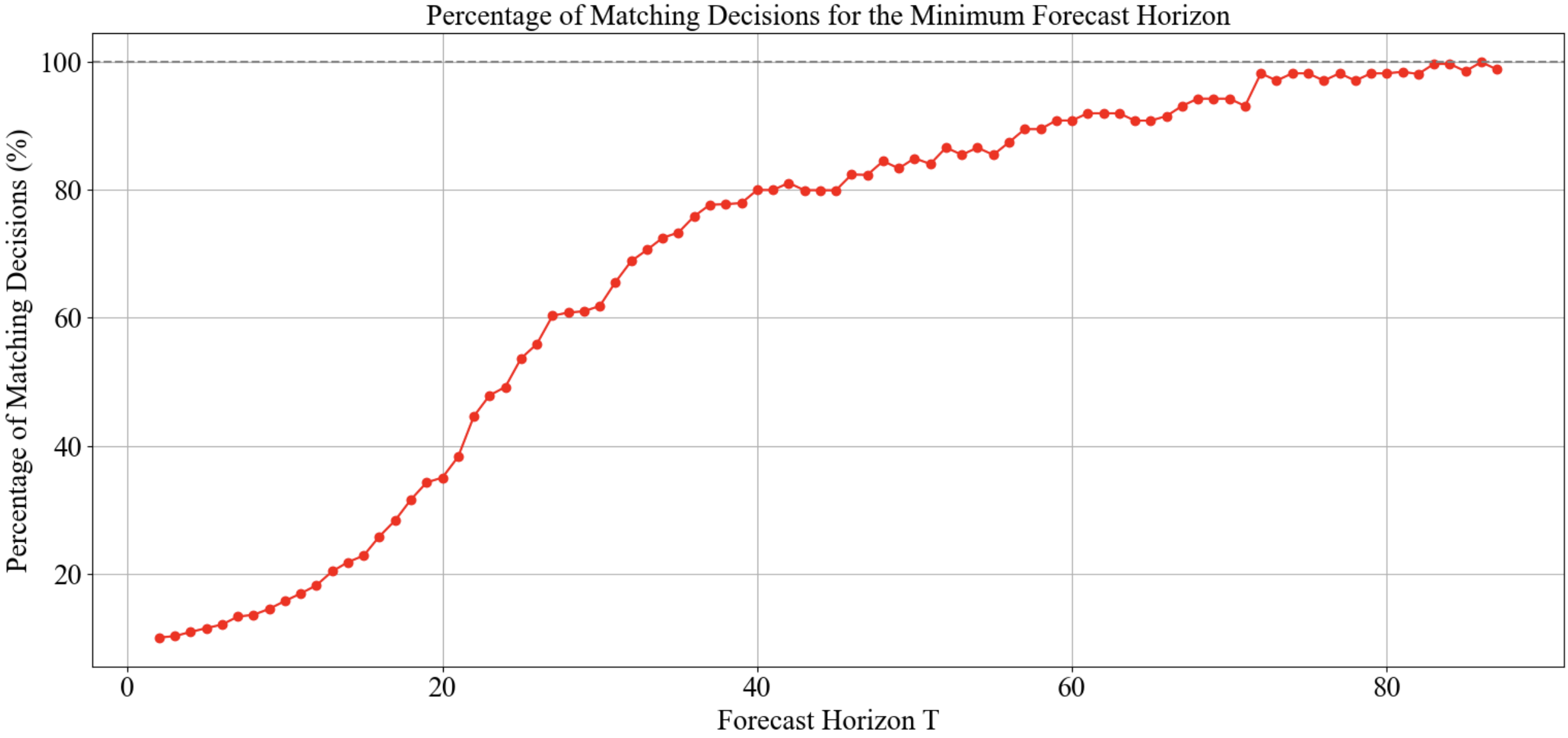}
\caption{Percentage of matching decisions using updated ESS parameters.}
\label{fig:no_min_horizon_percentage}
\end{figure}

The results in Figs.~\ref{fig:no_min_horizon_matrix} and \ref{fig:no_min_horizon_percentage} show that no forecast horizon 
$T \leq 88$ was sufficient to produce full control alignment with the global solution. These results confirm an important conclusion: the minimum forecast horizon is not guaranteed, which could be contingent upon the parameters of the ESS. Given the setup for the fast ESS in \cite{prat2024longlongenoughfinitehorizon}, the high round-trip efficiency $(0.9,0.9)$, absence of leakage $(\rho=1.0)$, as well as a balanced power-to-energy ratio, allows for very large sets of near-optimal yet different control sequences. Consequently, even for identical final states, successive actions vary, thus violating our more stringent criterion of trajectory matching.

This finding shows that the theorem in \cite{prat2024longlongenoughfinitehorizon} is merely a sufficient, not necessary, condition. Our analysis allows for a higher resolution of control fidelity and establishes that highly flexible regimes (such as full efficiency and no leakage) may support multiple globally optimal but trajectory-divergent policies.

\section{Conclusion}
In this paper, we studied the minimum forecast horizon in ESS scheduling. Inspired by the limitations of current terminal-state-based criteria (like in \cite{prat2024longlongenoughfinitehorizon}), we proposed a novel, operationally meaningful definition based on control trajectory matching and presented a methodological, algorithmic-based solution, giving rise to an effective framework for the optimization of such strategies. Our method takes into account the shortest planning horizon of $T^*$, such that, using a rolling horizon strategy, all of the first-stage control decisions can replicate the global optimal solution, within some numerical tolerance.

By using extensive simulation based on actual price data for the DK1 electricity market and a dynamic optimization structure, we derived the following main findings:

\begin{itemize}
    \item A forecast horizon of $T^* = 60$ hours was needed to exactly replicate the global optimal control actions and profits in this case study.

    \item Short horizons (up to 24 hours or shorter) that were traditionally utilized in practice were demonstrated to cause noticeable losses in both control discrepancy and economic return.

    \item A minimum forecast horizon is not necessarily guaranteed to exist: it is dependent on the ESS parameters, and convergence at the trajectory level might never be attained even if terminal conditions match.

    \item Our definition provides a stronger and more informative measure of control consistency and allows for a clearer understanding of forecast accuracy versus decision quality trade-off.
\end{itemize}

Although the study used numerical simulations to determine the shortest forecasting horizon, these techniques can be computationally expensive and provide few theoretical assurances. A future research direction could investigate analytical methods capable of defining sufficient and/or necessary conditions for horizon adequacy. An aim is to go beyond experimental validation and to establish provable requirements for horizon choice, likely in terms of closed-form constraints or theory-derived bounds. This would greatly alleviate computation and allow real-time or near real-time use of ESS scheduling in price uncertainty scenarios. In particular, we aim to:

\begin{itemize}
\item Derive requirements that associate ESS parameters (such as efficiency, storage capacity, and ramp rate) with the horizon that is needed for forecasting.

\item Develop provable guarantees for when a chosen horizon $T$ provides globally consistent or near-optimal control.

\item Expand analysis to stochastic price models and strong control formulations to increase real-world relevance under uncertain market situations.
\end{itemize}

By formalizing these concepts, we aim to advance more scalable, theory-driven, and market-responsive strategies for ESS involvement in electricity markets.

\bibliographystyle{IEEEtran}
\bibliography{IEEEabrv}

\end{document}